\documentstyle[prl,aps,twocolumn,psfig]{revtex}

\begin{document}

\twocolumn[\hsize\textwidth\columnwidth\hsize\csname
@twocolumnfalse\endcsname

\draft

\title{Emergence of a confined state in a weakly bent wire}
\author{Er'el Granot \footnotemark}

\address{School of Physics and Astronomy, Raymond and Beverly Sackler Faculty of Exact Sciences, Tel-Aviv University, 69978 Tel Aviv, Israel}

\date{\today}
\maketitle

\begin{abstract}
\begin{quote}
\parbox{16 cm}{\small
In this paper we use a simple straightforward technique to
investigate the emergence of a bound state in a weakly bent wire.
We show that the bend behaves like an infinitely shallow potential
well, and in the limit of small bending angle ($\varphi \ll 1$)
and low energy the bend can be presented by a simple 1D delta
function potential: $V\left( x\right) =-\left(
2\sqrt{c_{b}}\varphi ^{2}\right) \delta \left( x\right) $ where
$c_{b}\cong2.1$}
\end{quote}
\end{abstract}


]

\narrowtext \footnotetext{erel.g@chilight.com}
 \noindent It was well known for decades that the electric transmission of a quantum wire
(and, in general, any waveguide) is strongly affected by the
wire's boundaries' topology. Nevertheless, since 1989 many
researchers have validated a surprising finding. Exner and
Seba\cite{exner} were the first to show that a smoothly curved
waveguide holds a confined eigenstate, whose energy is lower than
the waveguide's cut-off energy. This bound state exists even when
there is no change in the waveguide's width. Avishai and coworkers
\cite{avishai} have used an elegant variational proof not only to
show that a bound state exists in a broken wire, but also to
evaluated its eigenvalue in the limit of small bending angle.
Later on Goldstone and Jaffe \cite {gold_jaffe} generalized these
findings and proved that any wire of constant width with {\em any
bend} will support at least one bound state (provided the wire
eventually straightens)

While the presence of such bound states was well proven by many
authors\cite {sols}-\cite{lin} their existence is still a puzzling
problem.

Carini et al.\cite{carini} suggested a qualitative explanation as
to why bends (and of course bulges) produce an effective
attraction and therefore a bound state. By substituting in the
Schrödinger equation a trial wave function for the lowest bound
state the problem is reduced to a 1D one. They showed that in this
case the bend can be regarded as an attractive (it is always
negative) 1D potential. In 1D, such a potential always has a bound
state.

Their qualitative description holds only in the adiabatic
approximation, i.e., when the wire's curvature is always small and
slowly changing. In the extreme case, where the bend occurs at a
single point (like the one discussed by Avishai et al.
\cite{avishai}, see Fig.1), such reasoning (which cannot be
applied) would yield an effective potential well, whose length is
proportional to the bending angle, $\varphi $(note, that the
potential depth is almost independent of $\varphi $). For such a
potential well the lowest bound state eigen energy goes like
$-\varphi ^{2}$.

Similarly, according to Sols and Macucci \cite{sols}, the bend can
be regarded as a small resonator whose effective width is slightly
larger than that of the waveguide in which it is introduced. This
larger effective width accounts for the lower minimum energy for
propagation, while the effective length of the resonator is
proportional to the bend's angle. Such a simplified qualitative
description again predicts bound eigen energy, which is
proportional to $-\varphi ^{2}$. These evaluations, however
contradicts the result of ref.\cite{avishai}, in which it was
proven that the eigen energy is quartic with respect to $\varphi $
(for small bending angles) . The quartic dependence also appears
in mildly curved wire (see ref. \cite{duclos} and references
therein).

\begin{figure}
\psfig{figure=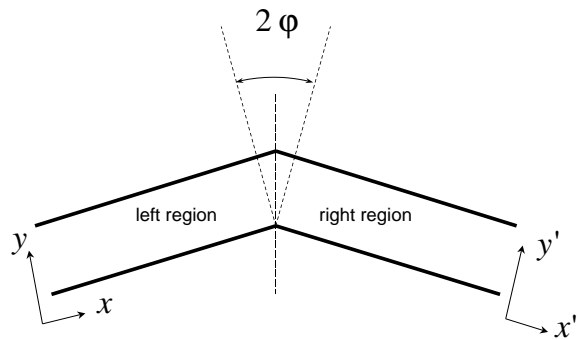,width=10cm,bbllx=8bp,bblly=23bp,bburx=820bp,bbury=578bp,clip=}
\caption{An illustration of the bent wire.}\label{fig1}
\end{figure}

The discrepancy appears since in the regime of weakly bending wire
(i.e., small bending angle) these simplified pictures (the cavity
picture, for example) cannot be applied. For example, in the
cavity picture most of the wavefunction will be distributed {\em
outside} the cavity.

In this short paper we show that in this case the bend behaves
like an infinitely shallow well(ISW), and that in the low energy
regime it can be replaced by a delta function potential in a 1D
wire \cite{delta}, and therefore cannot be presented by a simple
2D cavity.

\bigskip

There are several methods to calculate the bound-states eigen energies of a
bent wire (or waveguide). The most direct (and probably the most common)
method is to divide the wire into three parts: before the bend (a perfect
lead), the bent region, and after the bend (a perfect lead). This method is
usually used for a circular bend (or for one at right angles), and thus, the
complete set of wave functions is well known both in the bending region and
in the perfect leads. Thus, the solution is straightforward after matching
these solutions at the different regions' boundaries.

This method was carried out by Schult et al \cite{shult} to calculate the
eigen energies of an electron caught at the intersection of two narrow (but
totally diagonal) channels; Sols and Macucci \cite{sols} and Lent \cite{lent}
used this method to calculate the transmission through a circular bend. It
has also been used to calculate multiple bound states in a sharply bent
waveguide by Carini et al \cite{carini}, and in (long) circular bends by Lin
and Jaffe \cite{lin}.

A similar method is to use the Green function formalism: this approach was
taken by Goldstone and Jaffe \cite{gold_jaffe}, again for a right angle (but
rectangular) bend.

These two methods work very well for relatively large bending
angles. However, for low bending angles, they are not very
efficient, since they require extremely high accuracy of the
solutions of the Bessel functions. Since the difference between
the bound-state eigen energy and the cut-off (threshold) energy is
proportional to $\varphi ^{4}$ (\cite{avishai}, \cite{duclos}),
then even for $\varphi =10^{-2}$ (not to mention smaller angles),
the calculations require an accuracy (for the Bessel function
zeros) which is higher than $10^{-9}$.

A third popular method of calculating the bound states eigen energies is to
discretize the wave function over the entire 2D volume, and either by
iteration \cite{shult} or by a similar relaxation method \cite{carini}, to
calculate the low-lying eigenvalues and eigenfunctions of the problem.

Again, this method cannot be used in the small bending angle regime since
this would require extremely large matrices. For example, in the case $%
\varphi =10^{-2}$, the wave function will decay within a distance $10^{4}$
times larger than the wire's width. Therefore, in order to obtain the
required accuracy in the transversal direction (i.e., the wire's width), the
matrix will be too large to handle.

The best (and most elegant) method to calculate the eigen bound
state energy is the variational one (which was used in
ref.\cite{avishai}). However, this method cannot be applied to the
scattering case.

Hence, we use a slightly different and simpler approach. We do not
divide the wire into three regions but rather into only two, and
the only matching of the wave function is carried out at the
bending axis. Therefore, no Bessel functions are required, and
much higher accuracy can be achieved. Moreover, we match directly
the wave function, and therefore no overlapping integrals are
necessary, which makes it a very simple technique.

\bigskip

The geometry of the problem is illustrated in Fig.1. The stationary-state
Schr\"{o}dinger equation reads

\begin{equation}
\nabla ^{2}\psi \left( {\bf r}\right) +\left[ \omega -V\left( {\bf r}\right) %
\right] \psi \left( {\bf r}\right) =0  \label{shrodinger}
\end{equation}

(again we use the units $\hbar =2m=1$). $V$ is the potential of the wire's
walls (i.e., $V=0$ inside the wire and $V=\infty $ on the outside) and $%
\omega $ is the particle's energy.

Except for the bend, the system geometry is very simple; therefore, the
space can be divided into two regions: before the bend (say, left region)
and beyond it (say, right region). To simplify the notations, we use
different axes in each region: (x,y) and (x',y'), respectively (see Fig.1).

Should a bound state exists, it can be presented in the following way \cite
{granot}:

\begin{equation}
\begin{array}{lll}
\psi _{B}^{L}\left( x,y\right)&=\sum_{n=1}^{\infty }d_{n}\sin
\left(
k_{n}y\right) e^{\alpha _{n}x} & \text{\small for the left region} \\
\psi _{B}^{R}\left( x^{\prime },y^{\prime }\right)
&=\sum_{n=1}^{\infty }d_{n}\sin \left( k_{n}y^{\prime }\right) e^{
-\alpha _{n}x^{\prime }} & \text{\small for the right region}
\end{array}
 \label{row_solution}
\end{equation}

the subscript ''B'' represents bound state, and the superscripts ''L'' and
''R'' designate the left and right regions, respectively, and

\begin{equation}
k_{n}\equiv n\pi  \label{def_kn}
\end{equation}

and

\begin{equation}
\alpha _{n}\equiv \sqrt{\left( n\pi \right) ^{2}-\omega }.  \label{def_an}
\end{equation}

The strategy is the following \cite{granot}: $\psi _{B}^{L}$ is a solution
in the entire left region. We don't say {\em yet} that this is the right
one, but this is definitely a solution, because it solves the
Schr\"{o}dinger equation in the entire left region, and it agrees with the
boundary conditions of this region (except, for the moment, the one at $%
x\geq 0$). The same argument applies for $\psi _{B}^{R}$ : it solves the
Schr\"{o}dinger equation and maintains the boundary conditions in the entire
right region. Therefore it is a solution in that entire region.

Now, we need to find the right coefficients ($d_{n}$), which will
take care of the boundary condition at the break, i.e., the
continuity of the wave function and its derivative at the break
boundary. In order to do so, we match the wave function and its
derivative at $N$ different points on this line, then we take the
limit $N\rightarrow \infty $ and show that the solution (and the
coefficients) converge to a specific function.

Let us define a new set of coordinates:

\begin{equation}
\left(
\begin{array}{c}
\xi \\
\eta
\end{array}
\right) \equiv \left(
\begin{array}{cc}
\cos \varphi & \sin \varphi \\
-\sin \varphi & \cos \varphi
\end{array}
\right) \left(
\begin{array}{c}
x \\
y
\end{array}
\right) =\left(
\begin{array}{cc}
\cos \varphi & -\sin \varphi \\
\sin \varphi & \cos \varphi
\end{array}
\right) \left(
\begin{array}{c}
x^{\prime } \\
y^{\prime }
\end{array}
\right)   \label{coordinates}
\end{equation}

Then the wave function on the left side of the bend is:

\begin{equation}
\sum_{n=1}^{\infty }d_{n}\sin \left[ k_{n}\left( \xi \sin \varphi +\eta \cos
\varphi \right) \right] \exp \left[ \alpha _{n}\left( \xi \cos \varphi -\eta
\sin \varphi \right) \right]  \label{subst1}
\end{equation}

and on the right side:

\begin{equation}
\sum_{n=1}^{\infty }d_{n}\sin \left[ k_{n}\left( -\xi \sin \varphi +\eta
\cos \varphi \right) \right] \exp \left[ -\alpha _{n}\left( \xi \cos \varphi
+\eta \sin \varphi \right) \right]  \label{subst2}
\end{equation}

With these notations in mind, the matching of the wave function should take
place at $\xi =0$.

Limiting the calculations to $N$ modes, gluing of the wave function
derivative at $\xi =0$, i.e. requiring that $\partial \psi _{B}/\partial \xi
|_{\xi =0}=0$, leads to a single equation with $N$ variables. To solve them,
we quantize $\eta $ \cite{granot}:

\begin{equation}
\eta _{m}\equiv \frac{m-1}{\left( N-1\right) \cos \varphi }
\label{quant_eta}
\end{equation}

for $1\leq m\leq N$.

(these are the $N$ points were the matching takes place).

The prescribed substitution solves this problem: $N$ variables and $N$
equations, which can be written

\begin{equation}
\sum_{n=1}^{N}M_{nm}t_{n}=0  \label{equa_set}
\end{equation}

where

\begin{equation}
M_{nm}\equiv -\left[ M_{nm}^{1}+M_{nm}^{2}\right] \exp \left( -\alpha
_{n}\eta _{m}\sin \varphi \right) .
\end{equation}

$M_{nm}^{1}\equiv k_{n}\sin \varphi \cos \left( k_{n}\eta _{m}\cos \varphi
\right) $, and $M_{nm}^{2}\equiv \alpha _{n}\cos \varphi \sin \left(
k_{n}\eta _{m}\cos \varphi \right) $.

Clearly, a solution (a bound state) exists only when the matrix determinant
vanishes:

\begin{equation}
\left| M_{nm}\right| =0.  \label{deter}
\end{equation}

Solving eq.\ref{deter} numerically for $N\rightarrow \infty $\ one finds
that a confined solution exists and converges to

\begin{equation}
\omega _{b}\rightarrow \omega _{0}\equiv \pi ^{2}-c_{b}\varphi ^{4}
\label{sol_limit}
\end{equation}

where the proportionality constant converges to the theoretical
value \cite{avishai}

\begin{equation}
c_{b}\rightarrow 2.10\text{...}  \label{univ_const}
\end{equation}

Now, if our assumption is correct, and the bend can be presented
as an ISW in a 1D system in the limits $\varphi \rightarrow 0$ and
$\omega -\pi ^{2}\rightarrow 0$ then, it can be replaced by the
following 1D point potential \cite{delta} (in a 1D wire)

\begin{equation}
V\left( x\right) =-\left( 2\sqrt{c_{b}}\varphi ^{2}\right) \delta \left(
x\right)  \label{1D_potential}
\end{equation}

and the 1D transmission is obtained in straightforward fashion:

\begin{equation}
t=\frac{1}{1-i\sqrt{c_{b}}\varphi ^{2}/\Delta }
\label{1D_transmission}
\end{equation}

where

\[
\Delta \equiv \sqrt{\omega -\pi ^{2.}}
\]

In order to show that this 1D approximation is accurate for the limits $%
\varphi ,\Delta \rightarrow 0$, we will evaluate the transmission
in the direct approach.

Assume that the incident wave from $x=-\infty $ \ is

\begin{equation}
\psi _{inc}\left( x,y\right) =\sum_{n=1}^{\infty }\sin \left( k_{n}y\right)
\left[ a_{n}\exp \left( i\widetilde{k}_{n}x\right) +r_{n}\exp \left( -i%
\widetilde{k}_{n}x\right) \right]  \label{incident_wave}
\end{equation}

while the transmitted one ($x^{\prime }\rightarrow \infty $) is

\begin{equation}
\psi _{tran}\left( x^{\prime },y^{\prime }\right) =\sum_{n=1}^{\infty
}t_{n}\sin \left( k_{n}y^{\prime }\right) \exp \left( i\widetilde{k}%
_{n}x^{\prime }\right)  \label{transmitted_wave}
\end{equation}

where $a_{n},r_{n}$ and $t_{n}$ are the incident, reflected and transmitted
coefficients respectively [note that if $a_{n}=\delta _{n1}$ then $t$ (eq.%
\ref{1D_transmission})$\simeq t_{1}($eq.\ref{transmitted_wave}$)$]; $k_{n}\equiv n\pi ,$ and $\widetilde{k}_{n}\equiv \sqrt{%
\omega -\left( n\pi \right) ^{2}}$(i.e., $i\widetilde{k}_{n}=i\alpha _{n}$).

After transforming to the new coordinates (\ref{coordinates}) and solving by
following ref.\cite{granot} we obtain the transmission coefficient as a
function of the bending angle for a given set of coefficients $a_{n}$.

\begin{figure}
\psfig{figure=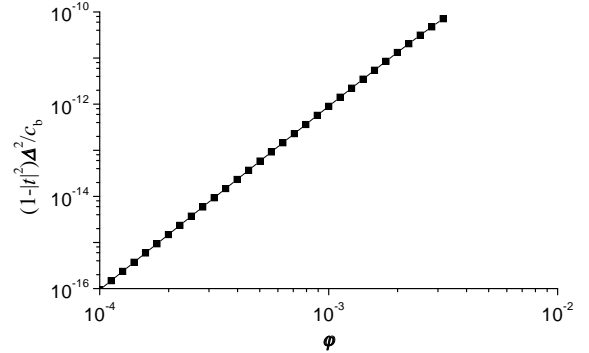,width=8cm,bbllx=-10bp,bblly=2bp,bburx=790bp,bbury=530bp,clip=}
\caption{Plot of the normalized reflection $\left( 1-\left|
t_{1}\right| ^{2}\right) \Delta ^{2}/c_{b}$ as a function of the
bend angle $\varphi $. This plot validates the approximation of
eq.\ref{small_angle_appr}.} \label{fig2}
\end{figure}

The plot of $1-\left| t_{1}\right| ^{2}$ as a function of the bend
angle is shown in Fig.3. As can be seen from this figure,
eq.\ref{1D_transmission}, which predicts

\begin{equation}
1-\left| t_{1}\right| ^{2}\sim c_{b}\varphi ^{4}/\Delta ^{2}
\label{small_angle_appr}
\end{equation}

(for small angles), is a very good approximation.

Hence, in the energy regime $c_{b}\varphi ^{4}/\Delta ^{2}\ll 1$,
the scattering wave function can be reduced to a 1D scattering
problem via the following separation of coordinates

\begin{equation}
\psi \left( \widetilde{x},\widetilde{y}\right) \simeq \sin \left( \pi
\widetilde{y}\right) \psi ^{1}\left( \widetilde{x}\right)  \label{appr_sol}
\end{equation}

where

\begin{equation}
\widetilde{x},\widetilde{y}=\left\{
\begin{array}{ccc}
x,y & \text{for} & x<0 \\
x^{\prime },y^{\prime } & \text{for} & x^{\prime }>0
\end{array}
\right\} \label{chan_coor}
\end{equation}

and $\psi ^{1}$ obeys the 1D Schr\"{o}dinger equation

\begin{equation}
-\frac{\partial \psi ^{1}}{\partial \widetilde{x}^{2}}-\left( 2\sqrt{c_{b}}%
\varphi ^{2}\right) \delta \left( \widetilde{x}\right) \psi
^{1}=\Delta ^{2}\psi ^{1}.  \label{1Dschrodingereq}
\end{equation}

Eq.\ref{appr_sol} also predicts the bound eigenstate with high accuracy:
\begin{equation}
\psi _{B}\simeq \sin \left( \pi \widetilde{y}\right) \exp \left( -\sqrt{c_{b}%
}\varphi ^{2}\left| \widetilde{x}\right| \right) .  \label{bound_state}
\end{equation}

Clearly, this approximation is accurate for $\left| \widetilde{x}\right|
\rightarrow \infty $.

\begin{figure}
\psfig{figure=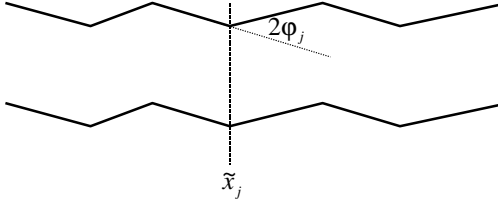,width=8cm,bbllx=20bp,bblly=23bp,bburx=720bp,bbury=578bp,clip=}
\caption{Wire with rough boundaries can be presented as a wire
with multiple bends} \label{fig3}
\end{figure}

Equation \ref{1Dschrodingereq} can easily be generalized to a wire
with an arbitrary number of bends (i.e., a rough boundaries wire,
see fig.3). Such a wire with rough boundaries can be presented by

\begin{equation}
-\frac{\partial \psi ^{1}}{\partial \widetilde{x}^{2}}-2\sqrt{c_{b}}%
\sum_{j=1}^{N}\varphi ^{2}_{j} \delta \left(
\widetilde{x}-\widetilde{x}_{j}\right) \psi ^{1}=\Delta ^{2}\psi
^{1}. \label{1Dschrodingerm}
\end{equation}

Before summarizing, it may be of interest to compare eq.\ref
{small_angle_appr}, i.e., the low energy scattering over the bend,
to scattering over a point impurity\cite{granot2}: when the
impurity is located a distance $\varepsilon $ from the wire's
boundary the wire's transmission should hold the relation

\begin{equation}
1-\left| t_{1}\right| ^{2}\sim c_{i}\varepsilon ^{4}/\Delta ^{2}
\label{impu}
\end{equation}

where the impurity's parameters are manifested in the coefficient $c_{i}$.

\bigskip

To summarise, in this paper we investigated the emergence of a
bound state in a bent wire. It was shown that in the limit of
small bending angle and low energy the system can be reduced to a
1D scattering problem, where the bend acts as a delta function
potential, i.e., $V\left( x\right) =-\left( 2\sqrt{c_{b}}\varphi
^{2}\right) \delta \left( x\right) $.

\bigskip

I would like to thank prof. Mark Azbel and prof. Yshai Avishai for
enlightening discussions.


\begin{references}
\bibitem{delta} If the width $W$ and depth $v$ of an arbitrary
well satisfy $|v|\ll W^{-2}$ the well is regarded as a shallow
one. Therefore, a delta function is an ideal ISW in 1D.
\end{references}
\end{document}